\documentclass[a4paper]{article}
\usepackage[english]{babel}
\usepackage[utf8]{inputenc}
\usepackage[T1]{fontenc}
\usepackage{amsmath}
\usepackage{graphicx}
\usepackage[colorinlistoftodos]{todonotes}
\usepackage[a4paper,left=2.5cm,right=2.5cm,top=2.cm,bottom=2cm]{geometry}
\usepackage{authblk}
\usepackage{txfonts}
\usepackage{mathptmx}
\usepackage{bbold} 
\usepackage{dsfont}
\usepackage{amsopn}  
\usepackage{amssymb}
\usepackage{graphicx}
\newtheorem{example}{Example}

\DeclareMathOperator*{\argmax}{arg\,max}

\title{A new quantum scheme for normal-form games}

\author{Piotr Fr\k{a}ckiewicz}
\affil{Institute of Mathematics\\ Pomeranian University, Poland}
\date{\today}

\begin{document}
\maketitle

\begin{abstract}
We give a strict mathematical description for a refinement of the Marinatto-Weber quantum game scheme. The model allows the players to choose projector operators that determine the state on which they perform their local operators. The game induced by the scheme generalizes finite strategic form game. In particular, it covers normal representations of extensive games, i.e., strategic games generated by extensive ones. We illustrate our idea with an example of extensive game and prove that rational choices in the classical game and its quantum counterpart may lead to significantly different outcomes.  
\end{abstract}
\section{Introduction}
\label{intro}
A 15-year-period research on quantum games results in many ideas of how a quantum game might look like and how it might be played. Certainly, the quantum scheme for $2\times 2$ games introduced in \cite{eisert} (the EWL scheme) has become one of the most common models and it has already found application in more complex games (see, for example, \cite{fracorpamiec}). However, the more complex a classical game is, the more sophisticated techniques are required to find optimal players' strategies in the EWL-type scheme. While in the scheme for $2\times 2$ games the result of the game depends on six real parameters (each players' strategy is a unitary operator from $\mathsf{SU}(2)$, and it is defined by three real parameters), the EWL-type scheme for $3\times 3$ games would already require 16 parameters to take into account \cite{restaurant}, \cite{su3}. One way to avoid cumbersome calculations when studying a game in the quantum domain was presented in \cite{marinatto} (see also recent papers \cite{bertrand}, \cite{khan} \cite{fracorrepeat} and \cite{situ} based on this scheme). The authors defined a model (the MW scheme) for quantum game where the players' unitary strategies were restricted to the identity and bit-flip operator. Then, the game became {\it quantum} if the players' local operators were performed on some fixed entangled state $|\Psi\rangle$ (called the players' joint strategy). The MW scheme appears to be much simpler than the EWL scheme. The number of pure strategies of each player is the same as in the classical game \cite{fracormarinatto}. Thus, the complexity of finding a rational solution is similar in both a classical game and the corresponding quantum counterpart. Unfortunately, that simple scheme exhibits some undesirable properties that we pointed out in \cite{fracormw}. First, the MW scheme implies {\it non-classical} game even if the players' joint strategy is an unentangled state. In particular, if a player's qubit is in an equal superposition of computational basis states, she cannot affect the game outcome in contrast to her strategic position in the classical game. Moreover, the players have no impact on the form of the initial state.
In paper \cite{fracormw} we showed that the above-mentioned drawbacks vanish by allowing the players to choose between the basis state that represents the classical game and the state $|\Psi\rangle$. In this paper, we continue that line of research. We give a formal description for players' strategies to include the choice of the initial state in the MW scheme. It will allow us to move beyond bimatrix games examined in \cite{fracormw} and consider more general normal form games. Then we study possible applications of the scheme. 

Some knowledge of game theory is required to follow this paper. While theory of bimatrix games is commonly used in quantum game theory, the notion of normal representation of extensive games may not be known for readers that deal with quantum games. Therefore, we encourage the reader who is not familiar with extensive game theory to see one of the textbooks~\cite{mejerson}, \cite{maszler}.
\section{Refinement of the Marinatto-Weber scheme}
In paper \cite{fracormw} we introduced a new scheme for playing finite bimatrix games in the quantum domain. The idea behind the scheme is that the players can choose whether they play a classical game or its quantum counterpart defined by the MW scheme. In the case of quantum model for $2\times 2$ bimatrix games, this means that the players choose their local operations: the identity $\mathds{1}$ or the Pauli operator $\sigma_{x}$ and additionally they decide whether the chosen operators are performed on state $|00\rangle$ or some fixed state $|\Psi\rangle \in \mathds{C}^2\otimes \mathds{C}^2$. Now, we give a formal description for the scheme. 
\label{sec:1}
\subsection{Quantum model for $2\times 2$ bimatrix game}
\label{subsec:1}
Let us consider a $2\times 2$ game
\begin{equation}\label{2x2}
\begin{pmatrix} (a_{00},b_{00}) & (a_{01}, b_{01}) \cr 
(a_{10}, b_{10}) & (a_{11}, b_{11}) \end{pmatrix}, ~~\mbox{where}~~(a_{ij},b_{ij})\in \mathds{R}^2.
\end{equation} 
The quantum scheme for game~(\ref{2x2}) is defined on an inner product space $(\mathds{C}^2)^{\otimes 4}$ by the following components:
\begin{enumerate}
\item A positive operator $H$,
\begin{equation}\label{positiveoperator1}
H = (\mathds{1}\otimes \mathds{1} - |11\rangle \langle 11|)\otimes |00\rangle \langle 00| + |11\rangle \langle 11|\otimes |\Psi\rangle \langle \Psi|,
\end{equation}
where $|\Psi\rangle\in \mathds{C}^2\otimes \mathds{C}^2$ such that $\| |\Psi \rangle \| = 1$, \item Players' pure strategies: $P^{(1)}_{i}\otimes U^{(3)}_{j}$ for player 1, $P^{(2)}_{k}\otimes U^{(4)}_{l}$ for player 2, where $i,j,k,l =0,1$, and the upper indices identify the subspace $\mathds{C}^2$ of $(\mathds{C}^2)^{\otimes 4}$ on which the operators
\begin{equation}\label{operatory}
P_{0} = |0\rangle \langle 0|,~ P_{1} = |1\rangle \langle 1|, \quad U_{0} = \mathds{1},~ U_{1} = \sigma_{x},
\end{equation}
are defined. That is, player 1 acts on the first and third qubit, player 2 acts on the second and fourth one. The order of qubits is in line with the upper indices.
\item Measurement operators $M_{1}$ and $M_{2}$ given by formula 
\begin{equation}\label{payoffmeasurement}
M_{1(2)} = \mathds{1}\otimes \mathds{1}\otimes \left(\sum_{x,y = 0,1}a_{xy}(b_{xy})|xy\rangle \langle xy|\right),
\end{equation}
where~$a_{xy}$ and $b_{xy}$ are the payoffs from (\ref{2x2}).
\end{enumerate}
The scheme proceeds in the similar way as the MW scheme or the EWL scheme---the players determine the final state by choosing their strategies and acting on operator~$H$. As a result, they determine the following density operator:
\begin{align}\label{pierwszydensity}
\rho_{\mathrm{f}} &= \left(P^{(1)}_{i}\otimes P^{(2)}_{k}\otimes U^{(3)}_{j}\otimes U^{(4)}_{l}\right)H \left(P^{(1)}_{i}\otimes P^{(2)}_{k}\otimes U^{(3)}_{j}\otimes U^{(4)}_{l}\right)\nonumber \\ &=\begin{cases} |11\rangle \langle 11|\otimes \left(U^{(3)}_{j}\otimes U^{(4)}_{l}|\Psi\rangle \langle \Psi| U^{(3)}_{j}\otimes U^{(4)}_{l}\right) &\mbox{if}~~i=k=1 \cr |ik\rangle \langle ik|\otimes \left(U^{(3)}_{j}\otimes U^{(4)}_{l}|00\rangle \langle 00|U^{(3)}_{j}\otimes U^{(4)}_{l}\right) &\mbox{otherwise}.  \end{cases}
\end{align}
Next, the payoffs for player 1 and 2 are
\begin{equation}
\mathrm{tr}(\rho_{\mathrm{f}}M_{1})~~\mbox{and}~~\mathrm{tr}(\rho_{\mathrm{f}}M_{2}).
\end{equation}
Similar to the MW scheme, each player is allowed to use mixed strategies, i.e., to choose her own strategies according to some probability distribution. Let $(p_{ij})_{ij=0,1}$ be a probability distribution over the set $\left\{P^{(1)}_{i} \otimes U^{(3)}_{j}\colon i,j =0,1\right\}$, and $(q_{kl})_{k,l = 0,1}$ be a probability distribution over $\left\{P^{(2)}_{k}\otimes U^{(4)}_{l}\colon k,l = 0,1\right\}$. Then the resulting density operator takes the form
\begin{equation}\label{mixeddensity}
\rho_{\mathrm{f}} = \sum_{\makebox[0pt]{$\scriptstyle i,j,k,l = 0,1$}}p_{ij}q_{kl}\left(P^{(1)}_{i}\otimes P^{(2)}_{k}\otimes U^{(3)}_{j} \otimes U^{(4)}_{l}\right)H\left(P^{(1)}_{i}\otimes P^{(2)}_{k}\otimes U^{(3)}_{j} \otimes U^{(4)}_{l}\right).
\end{equation}
Note that scheme (\ref{positiveoperator1})-(\ref{payoffmeasurement}) generalizes the classical way of playing the game. If the players' strategy profile takes the form
\begin{equation}
P^{(1)}_{0}\otimes P^{(2)}_{0} \otimes U^{(3)}_{j} \otimes U^{(4)}_{l},
\end{equation}
the players' payoffs depend on $U^{(3)}_{j}$ and $U^{(4)}_{l}$ and are equal to
\begin{equation}
\mathrm{tr}\left(\left(U^{(3)}_{j} \otimes U^{(4)}_{l}|00\rangle \langle 00| U^{(3)}_{j} \otimes U^{(4)}_{l}\right)\sum_{x,y = 0,1}a_{xy}(b_{xy})|xy\rangle \langle xy|\right) = a_{jl}(b_{jl}).
\end{equation}
Obviously, if $U^{(3)}_{j}$ and $U^{(4)}_{j}$ are chosen according to some probability distributions $\{p_{00}, p_{01}\}$ and $\{q_{00}, q_{01}\}$, respectively, the resulting distribution over $a_{jl}(b_{jl})$ coincides with one given by the corresponding mixed strategy profile in game~(\ref{2x2}). As a result, scheme (\ref{positiveoperator1})-(\ref{payoffmeasurement}) determines a game that is a complete quantization of~(\ref{2x2}) (see \cite{B} for the definition of complete quantization).

\paragraph{Nash equilibrium} In non-cooperative quantum game theory, Nash equilibrium is the most used solution concept. It is defined as a profile of strategies of all players in which each strategy is a best response to the other strategies. In view of scheme (\ref{positiveoperator1})-(\ref{payoffmeasurement}), it is a mixed strategy profile $\left((p^*_{ij})_{i,j=0,1},(q^*_{kl})_{i,j=0,1}\right)$ that solves the following optimization problems:
\begin{align}
&(p^*_{ij})\in\argmax_{(p_{ij})}\mathrm{tr}\left(\sum_{i,j,k,l = 0,1}p_{ij}q^*_{kl}S_{ikjl}HS_{ikjl}M_{1}\right)\label{1condition},\\
&(q^*_{kl})\in\argmax_{(q_{kl})}\mathrm{tr}\left(\sum_{i,j,k,l = 0,1}p^*_{ij}q_{kl}S_{ikjl}HS_{ikjl}M_{2}\right),\label{2condition}
\end{align}
where $S_{ikjl} = P^{(1)}_{i}\otimes P^{(2)}_{k}\otimes U^{(3)}_{j} \otimes U^{(4)}_{l}$. Like in the classical game theory, we can simplify conditions~(\ref{1condition}) and (\ref{2condition}) and only check if $(p^*_{ij})$ or $(q^*_{kl})$ yields a payoff that is equal to a maximum payoff when choosing pure strategies. More formally, condition~(\ref{1condition}) is equivalent to the following one
\begin{eqnarray}\label{equivcondition}
\mathrm{tr}\left(\sum_{i,j,k,l = 0,1}p^*_{ij}q^*_{kl}S_{ikjl}HS_{ikjl}M_{1}\right) = \max_{i,j = 0,1}{\mathrm{tr}\left(\sum_{k,l = 0,1}q^*_{kl}S_{ikjl}HS_{ikjl}M_{1}\right)}.\end{eqnarray}
It follows from the fact that $\mathrm{tr}(\rho_{\mathrm{f}}M_{1})$ for density operator $\rho_{\mathrm{f}}$ given by~(\ref{mixeddensity}) is a convex combination of elements
\begin{equation}
\mathrm{tr}\left(\sum_{k,l = 0,1}q^*_{kl}S_{ikjl}HS_{ikjl}M_{1}\right)~~\mbox{for}~~i,j =0,1
\end{equation}
with weights $p_{ij}$. In similar way we can simplify condition~(\ref{2condition}).
\paragraph{Bimatrix form} The game given by scheme (\ref{positiveoperator1})-(\ref{payoffmeasurement}) can be expressed  in terms of bimatrix form. Each entry of the bimatrix is a pair $\left(\mathrm{tr}(\rho_{\mathrm{f}}M_{1}), \mathrm{tr}(\rho_{\mathrm{f}}M_{2})\right)$ of payoffs that corresponds to a particular profile $P^{(1)}_{i}\otimes P^{(2)}_{k}\otimes U^{(3)}_{j} \otimes U^{(4)}_{l}$. As a result, we obtain
\begin{equation}\label{bimatrix}
\bordermatrix{&P^{(2)}_{0} \otimes U^{(4)}_{0} & P^{(2)}_{0} \otimes U^{(4)}_{1} & P^{(2)}_{1}\otimes
U^{(4)}_{0} & P^{(2)}_{1} \otimes U^{(4)}_{1} \cr  P^{(1)}_{0} \otimes
U^{(3)}_{0}&(a_{00},b_{00}) & (a_{01}, b_{01}) & (a_{00},b_{00}) &
(a_{01},b_{01})\cr P^{(1)}_{0}\otimes U^{(3)}_{1}&(a_{10},b_{10}) & (a_{11},
b_{11}) & (a_{10},b_{10}) & (a_{11},b_{11}) \cr
 P^{(1)}_{1}\otimes U^{(3)}_{0}&(a_{00},b_{00}) & (a_{01}, b_{01}) &
(\alpha_{00},\beta_{00}) & (\alpha_{01},\beta_{01}) \cr P^{(1)}_{1} \otimes
U^{(3)}_{1}&(a_{10},b_{10}) & (a_{11}, b_{11}) &
(\alpha_{10},\beta_{10}) & (\alpha_{11},\beta_{11}) }, \end{equation}
where 
\begin{equation}
(\alpha_{ij},\beta_{ij}) = \left(\mathrm{tr}(\rho_{ij}M_{1}), \mathrm{tr}(\rho_{ij}M_{2})\right)~~\mbox{for}~~\rho_{ij} = |11\rangle \langle 11|\otimes \left(U_{i}\otimes U_{j} |\Psi\rangle \langle \Psi| U_{i}\otimes U_{j}\right).
\end{equation}
Bimatrix (\ref{bimatrix}) is a very convenient way to study the game determined by scheme (\ref{positiveoperator1})-(\ref{payoffmeasurement}). Once, the entries $\left(\mathrm{tr}(\rho_{\mathrm{f}}M_{1}), \mathrm{tr}(\rho_{\mathrm{f}}M_{2})\right)$ are specified, we can leave quantum formalism out and use (\ref{bimatrix}). This is due to the linearity of trace that makes a density operator (\ref{mixeddensity}) and the corresponding probability distribution over pure strategies equivalent in a sense of generated outcomes. For example, in order to find Nash equilibria we can use the techniques for bimatrix games instead of conditions (\ref{1condition}) and (\ref{2condition}).

Note that bimatrix~(\ref{bimatrix}) clearly shows the role of components $P_{i}$ of players' strategies. Namely, the operations $U^{(3)}_{j}\otimes U^{(4)}_{l}$ are preformed on state $|\Psi\rangle$ if and only if both players form profile $P^{(1)}_{1}\otimes P^{(2)}_{1} \otimes U^{(3)}_{j}\otimes U^{(4)}_{l}$.

The scheme can be generalized to include more than one joint strategy $|\Psi\rangle$. Let us define operator $H$ on $\left(\mathds{C}^n\otimes \mathds{C}^{n}\right)\otimes \left(\mathds{C}^2\otimes \mathds{C}^2\right)$,
\begin{equation}
H = \left(\mathds{1}_{n^2\times n^2} - \sum^n_{i=1}|ii\rangle \langle ii|\right)\otimes |00\rangle \langle 00| + \sum^n_{i=1}|ii\rangle \langle ii|\otimes |\Psi_{i}\rangle \langle \Psi_{i}|
\end{equation}
and players' pure strategies
\begin{equation}
P^{(1)}_{i}\otimes U^{(3)}_{j}, P^{(2)}_{k}\otimes U^{(4)}_{l} \in \{|0\rangle \langle 0|, |1\rangle \langle 1|, \dots, |n\rangle \langle n|\}\otimes \{\mathds{1}, \sigma_{x}\}.
\end{equation}
In this case, the local operators $U^{(3)}_{j}\otimes U^{(4)}_{l}$ are performed on state $|\Psi_{i}\rangle$ if and only if the resulting stategy profile takes the form  $|ii\rangle \langle ii|\otimes U^{(3)}_{j}\otimes U^{(4)}_{l}$. 
\subsection{Quantum model for general bimatrix games}
We showed in \cite{fracormw2} how to construct the scheme for any finite bimatrix game according to the MW model. The key elements of the scheme are appropriately defined operators for players. In the case of $(n+1)\times (m+1)$ bimatrix game, 
\begin{equation}\label{nxm}
\begin{pmatrix}(a_{00},b_{00}) & (a_{01}, b_{01}) & \cdots & (a_{0m}, b_{0m})\cr 
(a_{10}, b_{10}) & (a_{11}, b_{11}) & \cdots & (a_{1m}, b_{1m})\cr \vdots & \vdots & \ddots & \vdots \cr (a_{n0},b_{n0}) & (a_{n1}, b_{n1}) & \cdots & (a_{nm},b_{nm})\end{pmatrix}, ~~(a_{ij},b_{ij})\in \mathds{R}^2.
\end{equation} 
where $n,m\geq 1$, player 1 (player 2) has $n+1$ operators $U_{i}$ ($m+1$ operators $V_{j}$) defined on space $\mathds{C}^{n+1}$ ($\mathds{C}^{m+1}$) that act on basis states $\{|0\rangle, |1\rangle, \dots, |n\rangle\}$ ($\{|0\rangle, |1\rangle, \dots, |m\rangle\}$) as follows:
\begin{eqnarray}\label{duzooperatorow}
 &&U_{0}|i\rangle = |i\rangle, ~~U_{1}|i\rangle = |i+1 ~\mathrm{mod}~ n+1\rangle, ~\dots ~~U_{n}|i\rangle = |i+n ~\mathrm{mod}~ n+1\rangle; \\ \label{duzooperatorow2} &&V_{0}|i\rangle = |i\rangle, ~~V_{1}|i\rangle = |i+1 ~\mathrm{mod}~ m+1\rangle, ~\dots ~~V_{m}|i\rangle = |i+m ~\mathrm{mod}~ m+1\rangle.
\end{eqnarray}
In view of~(\ref{duzooperatorow}) and (\ref{duzooperatorow2}), scheme (\ref{positiveoperator1})-(\ref{payoffmeasurement}) can be generalized by the players' strategies
\begin{equation}
\{P_{0},P_{1}\}\otimes \{U_{0},U_{1},\dots,U_{n}\}~~\mbox{and}~~\{P_{0},P_{1}\}\otimes \{V_{0}, V_{1},\dots, V_{m}\}.
\end{equation}
and the positive operator having the same form as~(\ref{positiveoperator1}), but with the outer product operators $|00\rangle \langle 00|, |\Psi\rangle \langle \Psi|$ defined on $\mathds{C}^{n+1}\otimes \mathds{C}^{m+1}$. 
\section{Quantum approach to finite normal form games}
In the previous section we formalized the refinement of the MW scheme that was introduced in \cite{fracormw}. We obtained the scheme that can be applied to any finite bimatrix game. In this section, we construct a framework for general normal-form games. The term of normal-form game has two main meanings. One concerns a strategic game given a priori. It is defined by triple $(N,\{S_{i}\}_{i\in N}, \{u_{i}\}_{i\in N})$, where $N$ is a set of players and, for $i\in N$, components $S_{i}$ and $u_{i}$ are player $i$'s strategy set and payoff function, respectively.
The second meaning concerns a strategic game $(N,\{S_{i}\}_{i\in N}, \{u_{i}\}_{i\in N})$ that is generated by a game in extensive form. The strategic game obtained in this way is called the normal representation of the extensive game. In what follows, we extend the scheme~(\ref{positiveoperator1})-(\ref{payoffmeasurement}) to cover both cases. 
\subsection{Strategic-form game}
The difference between bimatrix games and finite strategic games is that more than two players (say $n$ players) are allowed in the latter case. Therefore, operator~(\ref{positiveoperator1}) has to be modified in such a way that it simply outputs a density operator after $n$ players' strategies act on it. 

For simplicity of our analysis we restrict our attention to $n$-person strategic games with each $S_{i}$ having two elements. The extension of scheme (\ref{positiveoperator1})-(\ref{payoffmeasurement}) is defined now on space $(\mathds{C}^2)^{\otimes n} \otimes (\mathds{C}^2)^{\otimes n}$ with the positive operator $H$,
\begin{equation}\label{uogolnionyoperator}
H = \left(\mathds{1}^{\otimes n} - (|1\rangle \langle 1|)^{\otimes n}\right)\otimes (|0\rangle \langle 0|)^{\otimes n} + (|1\rangle \langle 1|)^{\otimes n} \otimes |\Psi\rangle \langle \Psi|,
\end{equation}
where $|\Psi\rangle \in (\mathds{C}^{2})^{\otimes n}$, $\||\Psi\rangle \| = 1$.
Each player $i\in \{1,\dots, n\}$ has a strategy determined by~(\ref{operatory}) that acts on qubits $i$ and $n+i$, i.e., it is on the form $P^{(i)}_{j_{i}}\otimes U^{(n+i)}_{j_{n+i}},$ where $j_{i}, j_{n+i} = 0,1.$
As a result, a profile of players' strategies forms operator $\left(\bigotimes^n_{i=1}P^{(i)}_{j_{i}}\right)\otimes \left(\bigotimes^{n}_{i=1}U^{(n+i)}_{j_{n+i}}\right)$ that results in the following density operator:
\begin{align}
\rho_{\mathrm{f}} &= \left[\left(\bigotimes^n_{i=1}P^{(i)}_{j_{i}}\right)\otimes \left(\bigotimes^{n}_{i=1}U^{(n+i)}_{j_{n+i}}\right)\right]H\left[\left(\bigotimes^n_{i=1}P^{(i)}_{j_{i}}\right)\otimes \left(\bigotimes^{n}_{i=1}U^{(n+i)}_{j_{n+i}}\right)\right] \nonumber\\ 
&= \begin{cases}(|1\rangle \langle 1|)^{\otimes n} \otimes \left[\left(\bigotimes^{n}_{i=1}U^{(n+i)}_{j_{n+i}}\right)|\Psi\rangle \langle \Psi| \left(\bigotimes^{n}_{i=1}U^{(n+i)}_{j_{n+i}}\right)\right] &\mbox{if}~ j_{1},\dots, j_{n} = 1 \cr \bigotimes^{n}_{i=1}|j_{i}\rangle \langle j_{i}|\otimes \left[\left(\bigotimes^{n}_{i=1}U^{(n+i)}_{j_{n+i}}\right)(|0\rangle \langle 0|)^{\otimes n}\left(\bigotimes^{n}_{i=1}U^{(n+i)}_{j_{n+i}}\right)\right] & \mbox{otherwise}.\end{cases}
\end{align}
Finally, we define for each player $i$ the payoff measurement $M_{i}$,
\begin{equation}\label{payoffm}
M_{i} = \mathds{1}^{\otimes n}\otimes \left(\sum_{x_{1},\dots,x_{n}=0,1} a^i_{x_{1},\dots,x_{n}}|x_{1}\dots x_{n}\rangle \langle x_{1}\dots x_{n}|\right),
\end{equation}
where $a^i_{x_{1},\dots,x_{n}}$ is player $i$'s payoff in the classical game that corresponds to strategy profile consisting of $(x_{1} + 1)$th strategy of player 1, $(x_{2} + 1)$ strategy of player 2, \dots, $(x_{n} + 1)$ strategy of player $n$.
It is not difficult to check that scheme~(\ref{uogolnionyoperator})-(\ref{payoffm}) generalizes a $n$-person strategic game with two strategies for each player. If the joint strategy $|\Psi\rangle$ is not played, i.e., element $\left(\bigotimes^n_{i=1}P^{(i)}_{j_{i}}\right)$ of a strategy profile is not equal to $(|1\rangle \langle 1|)^{\otimes n}$, then
\begin{align}
&\mathrm{tr}{\left[\left(\bigotimes^{n}_{i=1}|j_{i}\rangle \langle j_{i}|\right) \otimes \left[\left(\bigotimes^n_{i=1}U^{(n+i)}_{j_{n+i}} \right)(|0\rangle \langle 0|)^{\otimes n}\left(\bigotimes^n_{i=1}U^{(n+i)}_{j_{n+i}} \right)\right]M_{i}\right]} \nonumber \\ & = \mathrm{tr}{\left[\left(\bigotimes^{n}_{i=1}|j_{i}\rangle \langle j_{i}|\right) \otimes \left(\bigotimes^{n}_{i=1}|j_{n+i}\rangle \langle j_{n+i}|\right) M_{i} \right]} = a_{j_{n+1, \dots, 2n}}. 
\end{align}
Thus, for strategic-form game $\left(N, \{S_{i}\}_{i\in N}, \{u_{i}\}_{i\in N}\right)$,
\begin{equation}\label{strategicgame}
N = \{1,\dots, n\},~~S_{i} = \left\{s^{(i)}_{0}, s^{(i)}_{1}\right\},~~u_{i}\left(s^{(1)}_{k_{1}},\dots, s^{(n)}_{k_{n}}\right) = a^i_{k_{1},\dots, k_{n}}.
\end{equation}
The game generated by scheme (\ref{uogolnionyoperator})-(\ref{payoffm}) is equivalent to game~(\ref{strategicgame}) if strategies $s^{(i)}_{0}$ and $s^{(i)}_{1}$ are identified, respectively, with $U^{(n+i)}_{0}$ and $U^{(n+i)}_{1}$ for each $i$. 
\begin{example}\label{example1}
\textup{Let us consider the three-person Prisoner's Dilemma that was studied in the quantum domain (via the EWL scheme) by Du {\it et al.} \cite{du}. In terms of matrices the game is defined as follows:}
\begin{align}
\begin{tabular}{c c c c}\label{prisoner}
 & & \mbox{P 3} &  \\ & \mbox{P 2} & & \mbox{P 2} \\ \mbox{P 1} & $\begin{pmatrix}(3,3,3) & (2,5,2) \cr (5,2,2) & (4,4,0)\end{pmatrix}$ & & $\begin{pmatrix}(2,2,5) & (0,4,4) \cr (4,0,4) & (1,1,1)\end{pmatrix}$. \\
\end{tabular}
\end{align}
\textup{Here, player 1 and 2 choose between the rows and the columns, respectively, whereas player~3 chooses between the matrices. We recall that the only Nash equilibrium in~(\ref{prisoner}) is a profile consisting of the players' second strategies. Thus, the most reasonable result of the game is $(1,1,1)$. Similar to the best-known 2-person Prisoner's Dilemma, the players would increase their payoffs if at least two of them played their first strategies. However, the first strategy cannot be played by a rational player since for each profile of the opponents' strategies this strategy always yields a worse payoff than the second strategy. In what follows, we apply scheme (\ref{uogolnionyoperator})-(\ref{payoffm}) to game~(\ref{prisoner}). According to the reasoning used immediately before Example~\ref{example1}, we identify each player's strategies in game~(\ref{prisoner}) with local operators $U_{0}$ and $U_{1}$. Moreover, let us assume that player $i$, $i=1,2,3$ acts on the system of $i$th and $(i+3)$th qubit. As a result, scheme (\ref{uogolnionyoperator})-(\ref{payoffm}) comes down to one defined on $(\mathds{C}^{2})^{\otimes 3} \otimes (\mathds{C}^2)^{\otimes 3}$ with the positive operator}
\begin{equation}\label{operatorprzyklad}
H = \left(\mathds{1}^{\otimes 3} - |111\rangle \langle 111|\right)\otimes |000\rangle \langle 000| + |111\rangle \langle 111| \otimes |\Psi\rangle \langle \Psi|,
\end{equation}
\textup{the player $i$'s strategy set}
\begin{equation}
\left\{P^{(i)}_{0}\otimes U^{(i+3)}_{0}, P^{(i)}_{0}\otimes U^{(i+3)}_{1}, P^{(i)}_{1}\otimes U^{(i+3)}_{0}, P^{(i)}_{1}\otimes U^{(i+3)}_{1}\right\},
\end{equation}
\textup{and the triple of payoff operators}
\begin{align}\label{payoffprzyklad}
(M_{1}, M_{2}, M_{3}) &=  \mathds{1}^{\otimes 3}\otimes \bigl[(3,3,3)|000\rangle \langle 000| + (2,2,5)|001\rangle \langle 001| + (2,5,2)|010\rangle \langle 010|\nonumber \\ & \quad + (0,4,4)|011\rangle \langle 011| + (5,2,2)|100\rangle \langle 100| + (4,0,4)|101\rangle \langle 101|\nonumber \\ & \quad + (4,4,0)|110\rangle \langle 110| + (1,1,1)|111\rangle \langle 111|\bigr].
\end{align}
\textup{Let us fix now the players' joint strategy $|\Psi\rangle$ as:}
\begin{equation}
|\Psi\rangle = \frac{1}{2}\left(|001\rangle + |010\rangle + |100\rangle + |111\rangle\right)
\end{equation}
\textup{and determine the resulting players' payoffs that correspond to profiles}
\begin{equation}\label{profile64}
\bigotimes^{3}_{k=1}P^{(k)}_{j_{k}}\otimes \bigotimes^6_{k=4}U^{(k)}_{j_{k}},~j_{k} \in \{0,1\}.
\end{equation}
\textup{Note that for fixed $\bigotimes^6_{k=4} U^{(k)}_{j_{k}}$ the value}
\begin{equation}
\mathrm{tr}{\left[\left(\bigotimes^{3}_{k=1}P^{(k)}_{j_{k}} \otimes \bigotimes^6_{k=4} U^{(k)}_{j_{k}}\right) H \left(\bigotimes^{3}_{k=1}P^{(k)}_{j_{k}} \otimes \bigotimes^6_{k=4} U^{(k)}_{j_{k}}\right) M_{i}\right]},~i=1,2,3
\end{equation}
\textup{is the same for each $\bigotimes^{3}_{k=1}P^{(k)}_{j_{k}} \ne |111\rangle \langle 111|$. Therefore the problem of determining all the 64 payoff profiles actually reduces to determining $64 - 6\cdot 8 = 16$ of them. For example,}
\begin{align}
&\left(P^{(1)}_{1}\otimes P^{(2)}_{0} \otimes P^{(3)}_{0} \otimes U^{(4)}_{0}\otimes U^{(5)}_{0} \otimes U^{(6)}_{1}\right)H\left(P^{(1)}_{1}\otimes P^{(2)}_{0} \otimes P^{(3)}_{0} \otimes U^{(4)}_{0}\otimes U^{(5)}_{0} \otimes U^{(6)}_{1}\right) \nonumber \\ &\quad = |100\rangle \langle 100|\otimes \left(U^{(4)}_{0} \otimes U^{(5)}_{0} \otimes U^{(6)}_{1}\right)|000\rangle \langle 000|\left(U^{(4)}_{0} \otimes U^{(5)}_{0} \otimes U^{(6)}_{1}\right) \nonumber \\ &\quad = |100\rangle \langle 100| \otimes |001\rangle \langle 001|.
\end{align}
\textup{Then,} 
\begin{equation}
\mathrm{tr}{\left(|100\rangle \langle 100| \otimes |001\rangle \langle 001|M_{i}\right)} = \begin{cases}2 &\mbox{if}~ i\in \{1,2\} \cr 5 &\mbox{if}~ i=3.\end{cases}
\end{equation}
\textup{Hence, we obtain the same payoffs if $P^{(1)}_{1}\otimes P^{(2)}_{0} \otimes P^{(3)}_{0}$ is replaced by $P^{(1)}_{j_{1}}\otimes P^{(2)}_{j_{2}} \otimes P^{(3)}_{j_{3}} \ne |111\rangle \langle 111|$. For case $P^{(1)}_{1}\otimes P^{(2)}_{1} \otimes P^{(3)}_{1}$, we have}
\begin{eqnarray}\label{ostatnilabel}
&&\left(P^{(1)}_{1}\otimes P^{(2)}_{1} \otimes P^{(3)}_{1} \otimes U^{(4)}_{0}\otimes U^{(5)}_{0} \otimes U^{(6)}_{1}\right)H\left(P^{(1)}_{1}\otimes P^{(2)}_{1} \otimes P^{(3)}_{1} \otimes U^{(4)}_{0}\otimes U^{(5)}_{0} \otimes U^{(6)}_{1}\right) \nonumber \\  &&\quad= |111\rangle \langle 111| \otimes \left(U^{(4)}_{0}\otimes U^{(5)}_{0} \otimes U^{(6)}_{1}\right)|\Psi\rangle \langle \Psi|\left(U^{(4)}_{0}\otimes U^{(5)}_{0} \otimes U^{(6)}_{1}\right) \nonumber \\ &&\quad= |111\rangle \langle 111|\otimes \left(|\Psi'\rangle \langle \Psi'|\right),
\end{eqnarray}
\textup{where $|\Psi'\rangle = (|000\rangle + |011\rangle + |101\rangle + |110\rangle)/2$.
State~(\ref{ostatnilabel}) implies the payoff}
\begin{equation}
\mathrm{tr}{\left[|111\rangle \langle 111|\otimes \left(U^{(4)}_{0}\otimes U^{(5)}_{0} \otimes U^{(6)}_{1}\right)|\Psi\rangle \langle \Psi|\left(U^{(4)}_{0}\otimes U^{(5)}_{0} \otimes U^{(6)}_{1}\right)M_{i}\right]} = \frac{11}{4}
\end{equation}
\textup{for each $i=1,2,3$. Having determined the payoffs associated with each strategy profile, we can describe the game given by scheme~(\ref{operatorprzyklad})-(\ref{payoffprzyklad}) with the use of four matrices}
\begin{eqnarray}
 P^{(3)}_{0}\otimes U^{(6)}_{0}\quad \bordermatrix{&P^{(2)}_{0} \otimes U_{0}^{(5)} & P^{(2)}_{0} \otimes U_{1}^{(5)} & P_{1}^{(2)} \otimes U_{0}^{(5)} & P_{1}^{(2)}\otimes
U_{1}^{(5)} \cr  P^{(1)}_{0} \otimes U_{0}^{(4)} &(3,3,3) & (2,5,2) & (3,3,3) &
(2,5,2) \cr P^{(1)}_{0}\otimes U^{(4)}_{1} & (5,2,2) & (4,4,0) & (5,2,2) & (4,4,0) \cr
 P^{(1)}_{1}\otimes U^{(4)}_{0}&(3,3,3) & (2,5,2) &
(3,3,3) & (2,5,2) \cr P^{(1)}_{1} \otimes
U^{(4)}_{1}&(5,2,2) & (4,4,0) &
(5,2,2) & (4,4,0)}\nonumber \end{eqnarray} \begin{eqnarray} P^{(3)}_{0}\otimes U^{(6)}_{1}\quad \bordermatrix{&P^{(2)}_{0} \otimes U_{0}^{(5)} & P^{(2)}_{0} \otimes U_{1}^{(5)} & P_{1}^{(2)} \otimes U_{0}^{(5)} & P_{1}^{(2)}\otimes
U_{1}^{(5)} \cr  P^{(1)}_{0} \otimes U_{0}^{(4)} &(2,2,5) & (0,4,4) & (2,2,5) &
(0,4,4) \cr P^{(1)}_{0}\otimes U^{(4)}_{1} & (4,0,4) & (1,1,1) & (4,0,4) & (1,1,1) \cr
 P^{(1)}_{1}\otimes U^{(4)}_{0}&(2,2,5) & (0,4,4) &
(2,2,5) & (0,4,4) \cr P^{(1)}_{1} \otimes
U^{(4)}_{1}&(4,0,4) & (1,1,1) &
(4,0,4) & (1,1,1)} \nonumber\end{eqnarray} \begin{eqnarray} P^{(3)}_{1}\otimes U^{(6)}_{0}\quad \bordermatrix{&P^{(2)}_{0} \otimes U_{0}^{(5)} & P^{(2)}_{0} \otimes U_{1}^{(5)} & P_{1}^{(2)} \otimes U_{0}^{(5)} & P_{1}^{(2)}\otimes
U_{1}^{(5)} \cr  P^{(1)}_{0} \otimes U_{0}^{(4)} &(3,3,3) & (2,5,2) & (3,3,3) &
(2,5,2) \cr P^{(1)}_{0}\otimes U^{(4)}_{1} & (5,2,2) & (4,4,0) & (5,2,2) & (4,4,0) \cr
 P^{(1)}_{1}\otimes U^{(4)}_{0}&(3,3,3) & (2,5,2) &
(\frac{5}{2},\frac{5}{2},\frac{5}{2}) & (\frac{11}{4},\frac{11}{4},\frac{11}{4}) \cr P^{(1)}_{1} \otimes
U^{(4)}_{1}&(5,2,2) & (4,4,0) &
(\frac{11}{4},\frac{11}{4},\frac{11}{4}) & (\frac{5}{2},\frac{5}{2},\frac{5}{2})} \nonumber \end{eqnarray} \begin{eqnarray} P^{(3)}_{1}\otimes U^{(6)}_{1}\quad \bordermatrix{&P^{(2)}_{0} \otimes U_{0}^{(5)} & P^{(2)}_{0} \otimes U_{1}^{(5)} & P_{1}^{(2)} \otimes U_{0}^{(5)} & P_{1}^{(2)}\otimes
U_{1}^{(5)} \cr  P^{(1)}_{0} \otimes U_{0}^{(4)} &(2,2,5) & (0,4,4) & (2,2,5) &
(0,4,4) \cr P^{(1)}_{0}\otimes U^{(4)}_{1} & (4,0,4) & (1,1,1) & (4,0,4) & (1,1,1) \cr
 P^{(1)}_{1}\otimes U^{(4)}_{0}&(2,2,5) & (0,4,4) &
(\frac{11}{4},\frac{11}{4},\frac{11}{4}) & (\frac{5}{2},\frac{5}{2},\frac{5}{2}) \cr P^{(1)}_{1} \otimes
U^{(4)}_{1}&(4,0,4) & (1,1,1) &
(\frac{5}{2},\frac{5}{2},\frac{5}{2}) & (\frac{11}{4},\frac{11}{4},\frac{11}{4})} \nonumber
\end{eqnarray}
\textup{We see from the matrix representation that there are two types of pure Nash equilibria. The first one corresponds to the unique equilibrium in game~(\ref{prisoner}), and it is generated by profiles}
\begin{equation}\label{profile36}
\bigotimes^3_{k=1}P^{(k)}_{j_{k}} \otimes \bigotimes^{6}_{k=4}
U^{(k)}_{1},~~\mbox{\textup{where}}~~(j_{1},j_{2},j_{3}) \in \{(0,0,0), (1,0,0), (0,1,0), (0,0,1)\}.
\end{equation}
\textup{Each profile of~(\ref{profile36}) is a Nash equilibrium since each player's unilateral deviation from the equilibrium strategy yields the payoff 0 or 1. It also follows from the construction of (\ref{uogolnionyoperator})-(\ref{payoffm}). Namely, if a player cannot cause the joint strategy $|\Psi\rangle$ to be played by changing her own strategy, the equilibrium analysis is restricted to studying the local operations on state $|000\rangle$. That, in turn, coincides with the problem of finding Nash equilibria in game~(\ref{prisoner}), and $ \bigotimes^{6}_{k=4}U^{(k)}_{1}$ is just the counterpart of the profile of the players' second strategies that forms the unique equilibrium in~(\ref{prisoner}).
However, in contrast to~(\ref{prisoner}), the quantum game has another equilibrium given by profile}
\begin{equation}\label{profilenastepny}
\bigotimes^3_{k=1}P^{(k)}_{1} \otimes \bigotimes^6_{k=4}U^{(k)}_{1}.
\end{equation}
\textup{Indeed, player 1 suffers a loss of at least 1/4 by unilaterally deviation from strategy $P^{(1)}_{1}\otimes U^{(4)}_{1}$ and the same occurs in the case of player 2 and 3. Profile (\ref{profilenastepny}) is more profitable than (\ref{profile36}) since it implies 11/4 for each player instead of 1. Thus, the players gain by making use of the joint strategy $|\Psi\rangle$, i.e., by playing $\bigotimes^3_{k=1}P^{(k)}_{1}$. }\end{example}
\subsection{Normal representation of extensive games}
Given an extensive form game, one can construct a representation of that game in the strategic (normal) form. The resulting strategic game and the given extensive game have the same set of players and the same set of strategies for each player. The payoff functions are determined by the payoffs generated by the strategies in the extensive game. The normal representation appears to be a very convenient way to study the extensive game. In particular, while we lose the sequential structure, we obtain the sufficient and easier form of the game to find all the Nash equilibria. 

In our earlier paper~\cite{fracornormal}, we introduced a quantum scheme for playing an extensive game by using its normal representation. Basing on the MW and EWL schemes, we assigned an action at each information set in an extensive game to a local operation on a particular qubit in the quantum game. As a result, a number of qubits on which each player was allowed to specify local operations was equal to the number of their information sets.  In what follows, we extend our idea to the refinement of the MW scheme. This means that in addition to multiple choice of $\mathds{1}$ and $\sigma_{x}$, the players specify the state on which they perform the local operators.

Let us modify~(\ref{uogolnionyoperator}) to cover the normal-form game determined by an extensive game with the set of players $\{1,2,\dots, k\}$ and $n$ information sets, $n \geq k$. The positive operator is now defined on $(\mathds{C}^2)^{\otimes k} \otimes (\mathds{C}^{2})^{\otimes n}$ by formula
\begin{equation}\label{operatorostatni}
H = \left(\mathds{1}^{\otimes k} - (|1\rangle \langle 1|)^{\otimes k} \right) \otimes (|0\rangle \langle 0|)^{\otimes n} + (|1\rangle \langle 1|)^{\otimes k}\otimes |\Psi\rangle \langle \Psi|,
\end{equation}
where $|\Psi\rangle \in (\mathds{C}^{2})^{\otimes n}$ and $\||\Psi\rangle\| = 1.$ Let $\xi\colon \{k+1,k+2,\dots, k+n\} \to \{1,2, \dots, k\}$ be a surjective map. We define player $i$'s set of strategies as follows 
\begin{equation}\label{strategyostatni}
\left\{P^{(i)}_{j_{i}} \otimes \bigotimes_{\makebox[0pt]{$\scriptstyle y\in \xi^{-1}(i)$}}U^{(y)}_{j_{y}}\colon j_{i}, j_{y} \in \{0,1\}, i=1,2, \dots, k\}\right\}, 
\end{equation}
where $P^{(i)}_{j_{i}}$ and $U^{(y)}_{j_{y}}$ are defined by~(\ref{operatory}). As an possible application of~(\ref{operatorostatni})-(\ref{strategyostatni}), let us consider the following example:
\begin{example}[Four-stage centipede game]
\textup{A centipede game is a 2-person extensive game in which the players move one after another for finitely many rounds. In some sense, it can be treated as an extensive counterpart of the Prisoner's Dilemma. While both players are able to obtain a high payoff, their rationality leads them to one of the worst outcomes. An example of a four-stage centipede game is shown in Fig.~\ref{figure1}. Each player has two information sets (in this case, they are represented by the nodes of the game tree) with two available actions at each of them. Each player can stop the game (action S) or continue the game (action C), giving the opportunity to the other player to make her choice. One way to learn how the game may end is by backward induction. If player 2 is to choose at her second information set, she certainly plays action $S$ since she obtains 5 instead of 4---the result of playing action $C$. Since players' rationality is common knowledge, player 1 knows that by playing $C$ at her second information she ends up with payoff 3. Thus, player 1 chooses $S$ that yields 4. Similar analysis shows that the players choose action $S$ at their first information sets. Consequently, the backward induction predicts outcome $(2,0)$.}
\begin{figure}[t]
\centering
\includegraphics[scale=0.68]{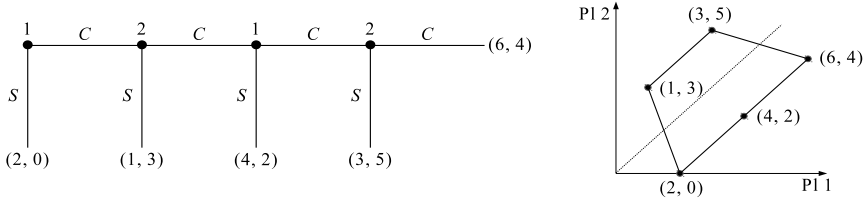}
\caption{Extensive form representation of a four-stage centipede game (left) and the corresponding payoff polytope (right).\label{figure1}}
\end{figure}
\textup{As we focus on normal form games, we construct the normal representation associated with the game in Fig.~\ref{figure1}. Let us first determine the players' strategies. We recall that a~player's strategy in an extensive game is a function that assigns an action to each information set of that player. Thus, each player has four strategies in the case of a four-stage centipede game. They can be written in the form $SS, SC, CS$ and $CC$, where, for example, $CS$ means that a player chooses $C$ at her first information set and $S$ at the second one. Once the strategies are specified, we determine the payoffs that correspond to all possible strategy profiles. For example, $(SC, CC)$ determines outcome $(2,0)$ since player 1's strategy $SC$ specifies action $S$ at her first information set. On the other hand, profile $(CC,CS)$ corresponds to payoff $(3,5)$ as player 1 always plays $C$ and player 2 chooses $S$ at her second information set. The players' strategies together with the payoffs corresponding to the strategy profiles define the following normal representation}
\begin{equation}\label{normalexample}
\bordermatrix{&SS & SC & CS & CC \cr  SS &(2,0) & (2,0) & (2,0) &
(2,0) \cr SC & (2,0) & (2,0) & (2,0) & (2,0) \cr
 CS &(1,3) & (1,3) &
(4,2) & (4,2) \cr CC &(1,3) & (1,3) &
(3,5) & (6,4)}.
\end{equation}
\textup{By using bimatrix~(\ref{normalexample}), we can learn that rational players always choose action $S$ at their first information sets. More formally, there are four pure Nash equilibria: $(SS,SS), (SS, SC), (SC, SS)$ and $(SC, SC)$, each resulting in outcome $(2,0)$.}

\textup{Let us consider the four-stage centipede game in terms of~(\ref{operatorostatni})-(\ref{strategyostatni}). We have $k=2$ and $n=4$. Thus, operator~(\ref{operatorostatni}) comes down to}
\begin{equation}\label{w1}
H = \left(\mathds{1}\otimes \mathds{1} - |11\rangle \langle 11|\right)\otimes |0000\rangle \langle 0000| + |11\rangle \langle 11| \otimes |\Psi\rangle \langle \Psi|.
\end{equation}
\textup{Let us assume that player 1 (player 2) performs her local operations on third and fifth (fourth and sixth) qubit, i.e., we define a map $\xi\colon \{3,4,5,6\} \to \{1,2\}$ by setting $\xi(\{3,5\}) = \{1\}$ and $\xi(\{4,6\}) = \{2\}$. According to~(\ref{strategyostatni}), player 1 and player 2's strategies take the form, respectively,}
\begin{equation}\label{w2}
P^{(1)}_{j_{1}} \otimes U^{(3)}_{j_{3}}\otimes U^{(5)}_{j_{5}}~~\mbox{\textup{and}}~~P^{(2)}_{j_{2}} \otimes U^{(4)}_{j_{4}}\otimes U^{(6)}_{j_{6}}~~\mbox{\textup{for}}~~j_{k}\in \{0,1\}.
\end{equation}
\textup{In order to generalize game~(\ref{normalexample}), we specify payoff operators (\ref{payoffm}) as follows}
\begin{align} 
(M_{1},M_{2}) &= \mathds{1}\otimes \mathds{1}\otimes \Biggl((2,0)\sum_{\makebox[0pt]{$\scriptstyle x_{2},x_{3},x_{4} \in \{0,1\}$}} |0x_{2}x_{3}x_{4}\rangle \langle 0x_{2}x_{3}x_{4}| + (1,3)\sum_{\makebox[0pt]{$\scriptstyle x_{3},x_{4} \in \{0,1\}$}}|10x_{3}x_{4}\rangle \langle 10x_{3}x_{4}|\nonumber\\  &\quad + (4,2)\sum_{\makebox[0pt]{$\scriptstyle x_{4} \in \{0,1\}$}}|110x_{4}\rangle \langle 110x_{4}| + (3,5)|1110\rangle \langle 1110| + (6,4)|1111\rangle \langle 1111| \Biggr). \nonumber
\end{align}
\textup{Setting $|\Psi\rangle = (|1010\rangle + |1011\rangle)/\sqrt{2}$ and determining}
\begin{equation}
\mathrm{tr}{\left[\left(P^{(1)}_{j_{1}}\otimes P^{(2)}_{j_{2}}\otimes \bigotimes^{6}_{k=3}U^{(k)}_{j_{k}}\right)H\left(P^{(1)}_{j_{1}} \otimes P^{(2)}_{j_{2}}\otimes \bigotimes^{6}_{k=3}U^{(k)}_{j_{k}}\right)M_{i}\right]}
\end{equation}
\textup{for player $i\in \{1,2\}$ and $j_{1},\dots, j_{6} \in \{0,1\}$, we obtain the following normal form game:}
\begin{equation}\label{ostatnimatrix}
\bordermatrix{&B_{000} & B_{001}& B_{010} & B_{011} &B_{100} & B_{101}& B_{110} & B_{111} \cr  A_{000} &(2,0) & (2,0) & (2,0) &
(2,0) &(2,0) &(2,0) &(2,0) &(2,0) \cr A_{001} & (2,0) & (2,0) & (2,0) & (2,0) &(2,0) &(2,0) &(2,0) &(2,0)\cr
 A_{010} &(1,3) & (1,3) &
(4,2) & (4,2) &(1,3) & (1,3) &
(4,2) & (4,2)\cr A_{011} &(1,3) & (1,3) &
(3,5) & (6,4) &(1,3) & (1,3) &
(3,5) & (6,4) \cr A_{100} &(2,0) &(2,0) &(2,0) &(2,0) & (1,3) & (1,3) & (\frac{9}{2},\frac{9}{2}) & (\frac{9}{2},\frac{9}{2}) \cr A_{101} & (2,0) & (2,0) & (2,0) & (2,0) & (1,3) & (1,3) & (4,2) & (4,2) \cr A_{110} & (1,3) & (1,3) & (4,2) & (4,2) &(2,0) &(2,0) &(2,0) &(2,0) \cr A_{111} & (1,3) & (1,3) & (3,5) & (6,4) &(2,0) &(2,0) &(2,0) &(2,0) },
\end{equation}
\textup{where $A_{j_{1}j_{3}j_{5}} = P^{(1)}_{j_{1}}\otimes U^{(3)}_{j_{3}} \otimes U^{(5)}_{j_{5}}$ and $B_{j_{2}j_{4}j_{6}} = P^{(2)}_{j_{2}}\otimes U^{(4)}_{j_{4}} \otimes U^{(6)}_{j_{6}}$. The game given by~(\ref{ostatnimatrix}) extends~(\ref{normalexample}) to local operations on $|\Psi\rangle \langle \Psi|$. If player 1 and 2 restrict their strategies, for example, to $P^{(1)}_{0} \otimes U^{(3)}_{j_{3}}\otimes U^{(5)}_{j_{5}}$ and $P^{(2)}_{0} \otimes U^{(4)}_{j_{4}}\otimes U^{(6)}_{j_{6}}$, $j_{3}, \dots, j_{6} \in \{0,1\}$, bimatrix~(\ref{ostatnimatrix}) boils down to~(\ref{normalexample}) (with the unique equilibrium outcome $(2,0)$). In general,  game~(\ref{ostatnimatrix}) has another Nash equilibrium}
\begin{equation}\label{paretoprofile}
(A_{100}, B_{110}) = P^{(1)}_{1}\otimes P^{(2)}_{1}\otimes U^{(3)}_{0}\otimes U^{(4)}_{1}\otimes U^{(5)}_{0} \otimes U^{(6)}_{0}
\end{equation}
\textup{that is not available in the classical game. Moreover, profile~(\ref{paretoprofile}) implies pair of payoffs $(9/2,9/2)$, that is the best possible symmetric outcome in~(\ref{normalexample}) (see, the payoff polytope in Fig.~\ref{figure1}). }
\end{example}
\textup{The main advantage of model (\ref{operatorostatni})-(\ref{strategyostatni}) or equivalently (\ref{uogolnionyoperator})-(\ref{payoffm}) is that a classical normal form game and its quantum counterpart have similar complexity. In particular, given any 2-person finite extensive game with $k$ strategies for each player, the normal form game implied by scheme (\ref{operatorostatni})-(\ref{strategyostatni}) is just a bimatrix $2k \times 2k$ game. As a result, there is no significant difference in the problem of determining Nash equilibria in both games.}
\begin{figure}[t]
\centering
\includegraphics[scale=0.69]{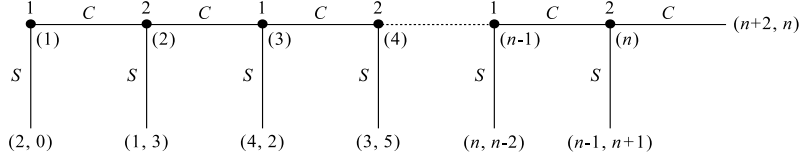}
\caption{$N$-stage centipede game.\label{figure2}}
\end{figure}
\begin{example}[$\mathbf{\textit{N}}$-stage centipede game] \textup{Let us consider a centipede game where this time the number of stages is any even integer $n$ for $n\geq 2$. The extensive form for this game is given in Fig.~\ref{figure2}. Similar to the four-stage centipede game, the $n$-stage case has also the unique equilibrium outcome $(2,0)$. Rational players choose action $S$ at their own information sets  even though the game enables the players to obtain the payoffs approximate to the number of stages. We have learned from the preceding example that there is a unique, symmetric, and pareto-optimal Nash equilibrium if (\ref{normalexample}) is extended to (\ref{ostatnimatrix}). It turns out that the result is valid in the general case. That is, there is a Nash equilibrium that implies the payoff $n + 1/2$ for both players (pair of payoffs $(n + 1/2, n+ 1/2)$ is indeed a paretooptimal outcome since it is the midpoint of the segment whose endpoints are $(n-1,n+1)$ and $(n+2,n)$).  In order to prove the existence of that equilibrium, let us generalize (\ref{w1}) and (\ref{w2}) to an arbitrary $n$-stage centipede game. Since there are two players and $n$ information sets in the game, the positive operator $H$ and the players' strategies are given by (\ref{operatorostatni}) and (\ref{strategyostatni}) for $k=2$. We assume that players 1 and 2 perfom their local operators on qubits with odd and even indices, respectively. Thus, the map $\xi\colon \{3,4,\dots, n+2\} \to \{1,2\}$ is given by formula}
\begin{equation}
\xi(x) = \begin{cases}1 &\mbox{if}~ x ~\mbox{\textup{is odd}} \cr 2 &\mbox{if}~ x ~\mbox{\textup{is even.}}\end{cases}
\end{equation}
\textup{The appropriately  generalized payoff operators take the form}
\begin{align}\label{payoffncentipede}
(M_{1},M_{2})&= \mathds{1}\otimes \mathds{1} \otimes \Biggl((2,0)\sum_{\makebox[0pt]{$\scriptstyle x_{2}\dots x_{n} \in \{0,1\}$}}|0x_{2}\dots x_{n}\rangle \langle 0x_{2}\dots x_{n}| \nonumber\\ &\quad + (1,3)\sum_{\makebox[0pt]{$\scriptstyle x_{3}\dots x_{n} \in \{0,1\}$}}|10x_{3}\dots x_{n}\rangle \langle 10x_{3}\dots x_{n}|\nonumber \\ &\quad + (4,2)\sum_{\makebox[0pt]{$\scriptstyle x_{4}\dots x_{n} \in \{0,1\}$}}|110x_{4}\dots x_{n}\rangle \langle 110x_{4}\dots x_{n}| + \dots  \nonumber \\ &\quad + (n,n-2)\sum_{\makebox[0pt]{$\scriptstyle x_{n} \in \{0,1\}$}}|11\dots 10x_{n}\rangle \langle 11\dots 10x_{n}| \nonumber \\ &\quad + (n-1,n+1)|11\dots 10\rangle \langle 11\dots 10| \nonumber \\ &\quad + (n+2,n)|11\dots 11\rangle \langle 11\dots 11|\Biggr).
\end{align}
\textup{Let us consider the state $|\Psi\rangle\in (\mathds{C}^{2})^{\otimes n}$,}
\begin{equation}
|\Psi\rangle = \frac{|1010\dots 1010\rangle + |1010 \dots 1011\rangle}{\sqrt{2}}
\end{equation}
\textup{and a strategy profile $U^* \otimes V^*$ such that}
\begin{equation}
U^{*} = P^{(1)}_{1}\otimes \bigotimes_{\makebox[0pt]{$\scriptstyle y \in \xi^{-1}(1)$}}U^{(y)}_{0}~~\mbox{\textup{and}}~~V^* = \left(P^{(2)}_{1} \otimes \bigotimes_{\makebox[0pt]{$\scriptstyle y \in \xi^{-1}(2), y \ne n+2$}}U^{(y)}_{1} \right)\otimes U^{(n+2)}_{0}.
\end{equation}
\textup{First note that strategy profile $U^*\otimes V^*$,}
\begin{equation}
U^*\otimes V^* = |11\rangle \langle 11| \otimes \mathds{1}^{(3)}\otimes \sigma^{(4)}_{x} \otimes \mathds{1}^{(5)} \otimes \sigma^{(6)}_{x} \otimes \dots \otimes \mathds{1}^{(n-1)} \otimes \sigma^{(n)}_{x} \otimes \mathds{1}^{(n+1)} \otimes \mathds{1}^{(n+2)}
\end{equation}
\textup{implies the payoffs}
\begin{eqnarray}\label{equilibriumpayoff}
\mathrm{tr}{\left[\left(U^*\otimes V^*\right)H\left(U^*\otimes V^*\right)M_{i}\right]} = n + \frac{1}{2}~~\mbox{\textup{for}}~~i = 1,2.
\end{eqnarray}
\textup{Let $U = P^{(1)}_{j_{1}}\otimes \bigotimes_{y\in \xi^{-1}(1)}U^{(y)}_{j_{y}}$ be an arbitrary player 1's strategy. If $j_{1} = 0$, then}
\begin{equation}
\left(U\otimes V^*\right)H\left(U\otimes V^*\right) = |01\rangle \langle 01|\otimes \left(U^{(3)}_{j_{3}} \otimes \dots \otimes U^{(n+1)}_{j_{n+1}}\right)|01\dots 0100\rangle \langle 01\dots 0100|\left(U^{(3)}_{j_{3}} \otimes \dots \otimes U^{(n+1)}_{j_{n+1}}\right).
\end{equation}
\textup{Since player 1 cannot affect the system of $(n+2)${\it th} qubit, we have }
\begin{equation}
\max_{\makebox[0pt]{$\scriptstyle \bigotimes_{y\in \xi^{-1}(1)}U^{(y)}_{j_{y}}$}}{\mathrm{tr}{\left[\left(U\otimes V^{*}\right)H\left(U\otimes V^{*}\right)M_{1}\right]}} = n < n + \frac{1}{2}.
\end{equation}
\textup{In the case of $j_{1} = 1,$ }
\begin{equation}
\left(U\otimes V^*\right)H\left(U\otimes V^*\right)= |11\rangle \langle 11|\otimes \left(U^{(3)}_{j_{3}} \otimes \dots \otimes U^{(n+1)}_{j_{n+1}}\right)|\varphi\rangle \langle \varphi|\left(U^{(3)}_{j_{3}} \otimes \dots \otimes U^{(n+1)}_{j_{n+1}}\right),
\end{equation}
\textup{where $|\varphi \rangle = (1/\sqrt{2})(|11\dots 10\rangle + |11\dots 1\rangle).$ From~(\ref{equilibriumpayoff}), we know that player 1 gets $n + 1/2$ if $U = U^*$. Thus, the form of (\ref{payoffncentipede}) implies that $U \ne U^*$ would increase the player 1' payoff only if $U$ made the magnitude of the amplitude of $|11\dots 1\rangle$ higher that $1/\sqrt{2}$. However, it is not possible because of the form of $U$. As a result, we have proved that $U^*$ is a player 1' best response to $V^*$ over all her pure strategies. Using a similar argument to one concerning the equivalence of~(\ref{1condition}) and (\ref{equivcondition}), we conclude that $U^*$ is a player 1' best response to $V^*$ over all her (pure and mixed) strategies. In similar way, we can show that player 2' strategy $V^*$ is a best response to $U^*$.}
\end{example}
\section{Conclusions}
The aim of our research was to formalize our idea about the MW-type schemes. As a result, we have showed that the players' strategies do not have to be unitary operators or even superoperators in the quantum game. Apart from unitary operators, they may include projectors that determine the state on which the unitary operations are performed. Thus, the initial state does not have to be a density operator. Certainly, the scheme is in accordance with the laws of quantum mechanics. The resulting state is given by a density operator, and therefore the payoff measurement is well-defined. A positive point of the scheme is a way it can be considered. Given a bimatrix game the scheme outputs a bimatrix game. Consequently, it implies similar complexity in finding optimal strategies for the players. In addition, our model enables us to consider extensive games via the normal representation. Moreover, the example of the general centipede game has proved that the analysis does not have to be limited to simple games. We suppose that this argument may attract the attention of researchers to the refinement of the MW scheme. 

\end{document}